\documentclass{kapproc} 
\usepackage{procps} 
\usepackage{amssymb}
\usepackage{latexsym} 
\usepackage{mathptm}
\usepackage[dvips]{graphicx}

\upperandlowercase
\nochapequationnumber 
\let\footnote\savefootnote
\let\footnotetext\savefootnotetext

\normallatexbib 

\begin{document}

\articletitle{Signatures of Large Extra Dimensions}

\author{S.~Hossenfelder, M.~Bleicher and H.~St\"ocker}
\affil{Institut f\"ur Theoretische Physik\\ J. W. Goethe Universit\"at\\
 D-60054, Frankfurt am Main, Germany}
\email{hossi@th.physik.uni-frankfurt.de}

\begin{abstract}
String theory suggests modifications of our spacetime such as extra dimensions
and the existence of a mininal length scale. In models with addidional 
dimensions, the Planck scale can be lowered to values accessible by future 
colliders. Effective theories which extend beyond the standart-model by including extra dimensions and a minimal length allow computation of observables and
can be used to make testable predictions. Expected effects that arise
within these models are the production of gravitons and black holes. Furthermore,
the Planck-length is a lower bound to the possible resolution of spacetime which might be reached soon.

This is a summary of a talk given at the 
{\sc NATO} Advanced Study Institute in Kemer, Turkey, Oct. 2003.
\end{abstract}


\section{Introduction}
The standard model (SM) of particle physics yields an extremely precise theory for 
the electroweak and strong interaction. It allowed us to improve our view of nature
in many ways but leaves us with several unsolved problems. E.g. we still have to understand 
the large number of free parameters in the SM, 
the question of the fermion families, the mechanism of electroweak symmetry breaking, CP violation,
and of course the puzzle of quantum gravity. 

Gravity, if it is quantized as a spin-2 field in the canonical way, is non-renormizable. The only
reason why  the SM yields such high accuracy without concerning gravity is
that gravity is much weaker than the other interactions. The effects of quantum gravity get
as important as the effects of the SM only at the so-called Planck-scale which is
reached at energies near the Planck-mass of $m_{\rm p} \approx 10^{16}$ TeV or at distances near the 
Planck-length $l_{\rm p}\approx 10^{-20}$~fm, resp. This comparable weakness of the gravitational
interaction, also known als the ''hierarchy-problem'', is another yet unexplained fact which singles out gravity and has to be
understood for achieving a successful unifying theory.

All these problems have in common that they can not be explained within the SM itself.
The SM, it seems, is a low energy limit of a more general theory. It is one of the most exciting and
challenging tasks on physicists in the 21st century to go on and look beyond 
the SM.  

On the one hand, huge steps in this direction have been undergone by string theorists in the 
last decades. Supersymmetric string theory is -- up to current knowledge -- todays most promising candidate of a 
Grand Unified Theory (GUT). In addition to the SM symmetries, it provides naturally the
existence of a spin-2 particle. Moreover, this canditate for quantum gravity is
finite (at least in pertubation theory). 

On the other hand, the need to look beyond infected many
experimental groups which search for SM - violating processes.

Unfortunately, there is a gap between theory and experiment as the most obvious 
predictions of string theory -- excited particles, one loop corrections -- have to be hidden by broken symmetry at low enery scales and thus, there is no way so far to verify string theory. 
Whenever there is a theory in its full mathematical 
beauty that can not be applied to computate observables, one needs to make approximations.
In a certain way, theoretical physics is the art of approximation; it is the art of model building; it is the art of simplification. 

\vspace{0.5cm}
{\sl ''Science may be described 
as the art of systematic over-simplification.''

\hspace{3cm} --- Karl Popper, The Observer, August 1982}
\vspace{0.5cm}

The recently proposed models of extra dimensions are models that can be used to fill 
the gap between theoretical conclusions and experimental possibilities. These models are motivated  by string theory but do not have to cope with all the 
stringy implications. They are kind of an effective model for a theory beyond the
SM. The main idea taken from string theory are the notions of open and closed strings which provide us naturally with two different
kinds of particles. The closed strings describe the graviton, the open
strings the other interacting particles. So, there is a natural reason why gravity is
different. Further, supersymmetric string theory does only work properly 
(no anomalies) in spacetimes with extra dimensions. These extra dimensions
are compactified to finite extension by empiric means: we have not seen them. When compactifying the
extra dimensions we have to confine the open strings to be attached on a three-dimensional 
submanifold: our 3-brane, the universe we are used to.
 
There are different ways to build a model of extra dimensional space-time. Here, we want to
mention only the most common ones:
\begin{enumerate}
\item The {\sc ADD}-model proposed by N. Arkani-Hamed, S. Dimopoulos and G. Dvali \cite{add} adds $d$ extra
spacelike dimensions without curvature, in general each of them compactified to the same radius $R$. All SM particles are confined to our brane, while gravitons are allowed to propagate freely in the bulk. 
\item The setting of the model from L.~Randall and R. Sundrum \cite{rs1,rs2} is a 5-dimensional spacetime with
an non-factorizable geometry. The solution for the metric is found by analyzing the solution of Einsteins field equations with an energy density on our brane, where the SM particles live. In the type I model \cite{rs1} the extra dimension is compactified, in the type II model \cite{rs2} it is infinite.
\item Within the model of universal extra dimensions \cite{uxds}
all particles (or in some extensions, only bosons) can propagate in the whole multi-dimen\-sional spacetime. The extra dimensions are compactified on an orbifold to reproduce stand\-ard model gauge degrees of freedom.
\end{enumerate}

In the following we will focus on the ADD-model. For a more general review the reader is refered to \cite{review}.

The {\sc ADD}-model explains the hierarchy between the electroweak and the Planck-scale with the large volume of the extra dimensions. Consider the Poisson-equation for
a point particle of mass $m$ in $d+3$ spacelike dimensions. The coupling constant will have dimension mass$^{-d-2}$. This new mass-scale is the new higher dimensional Planck-mass and will be denoted by $M_{\rm f}$, the new Planck-length is $L_{\rm f}=1/M_{\rm f}$. 

The power law for the
potential $V(r)$ goes with the distance $r$ from the source as $r^{-d-1}$. This holds in the compactified scenario for distances much smaller than the radius $R$ of the extra dimensions. For large distances $r\gg R$, $d$ of the powers factorize and we have to match this to the usual $1/r$ power law in three dimensions with the familiar coupling $m_{\rm p}^2$.
This yields
\begin{eqnarray}
\frac{V}{m} &=& \frac{1}{M_{\rm f}^{d+2}}\frac{1}{r^{d+1}} \to 
\frac{1}{M_{\rm f}^{d+2}}\frac{1}{R^{d}}\frac{1}{r} = \frac{1}{m_{\rm p}^{2}}\frac{1}{r}\quad,
\end{eqnarray}
and so one obtains the relation
\begin{eqnarray} \label{master}
m_p^2=M_{\rm f}^{2+d} R^d\quad.
\end{eqnarray}
With the assumption $M_{\rm f} \sim$TeV, the compactification scale $R$ ranges from\linebreak $1/10$~mm to $10^2$~fm, resp. $1/R$ from $10^{-2}$~eV to 10~MeV, if $d$ runs from $2$ to $7$. $d$=1 is excluded since this
would imply an extra dimension of the size of the solar system.
It was shown, that this setting with large\footnote{Large when compared to the Planck-scale.} extra dimensions ({\sc LXD}s) can be motivated by string theory and it indeed lowers the unification scale to values $\approx M_{\rm f}$ \cite{dienesundso}.

The lowered Planck-scale leads to a vast number of observable effects. The most
obvious one is a modification of Newtons law at small distances, which is in todays measurable range  for $d=2$. There are several groups working on this sub-mm gravity measurements \cite{submm}. So far,  $R>$0.18 mm can be excluded. Next generation experiments are expected to yield precise measurements up to $\mu$m-distances.

This increase of the gravitational force at small distances enables the production of black holes at energy scales $\approx M_{\rm f}$ which can be reached at the {\sc LHC}. Further, with the lowered scale, the production of gravitons becomes significant at energies $\approx M_{\rm f}$. Also, there are contributions to cross sections from the virtual graviton exchange. 
In the following, we will briefly discuss possible consequences for high energy physics, observable effects and the 
extension of the model to include minimal-scale effects.

\section{Gravitons}

Gravitons are treated as pertubations\footnote{Capital indices run from $0$ to $d+1$, small latin indices run over the extra dimensions $4..4+d$ only, small greek indices run over the non-compactified dimensions, from $ 0$ to $3$.} $h_{IJ}$ of the higher dimensional metric tensor $g_{IJ}=\eta_{IJ}+h_{IJ}$. For an effective description on our brane, this pertubation tensor can be decomposed in a spin-2 tensor, which describes the graviton, vector fields and scalar fields. From this ansatz one obtains the Lagrangian by minimal coupling to SM fields. The analysis shows that only the
spin-2 field couples to the energy-momentum tensor.\footnote{The trace of the scalar fields, also known as dilaton or radion, couples to the trace of the energy-momentum tensor.} With the interaction terms, one can then derive the Feynmann rules \cite{gravana} for the quantized fields. This enables us
to compute graviton cross sections at least at tree-level. 

In absence of matter fields, the fields obey an $d+3$ dimensional wave equation. 
Due to the periodicity of the extra dimensions, we can expand the solutions in a fourier series with $n_i/R$, where $n=(n_1,..,n_d)$ is the number-vector of the excitation level. The quantized momentum in the direction of the extra dimensions yields an apparent mass term for the graviton if described effectively on our brane: 
\begin{eqnarray*}  
\eta^{IJ} \partial_I \partial_J = \Box  - \; \sum_{d}\frac{n_i^2}{R^2} \quad. 
\end{eqnarray*}
Thus, we have a tower of massive gravitons. Since the level spacing $1/R$ is tiny compared to collider energies, the number $N(\sqrt{s})$ of excited graviton levels, that can be occupied with an energy $\sqrt{s}$ is $N(\sqrt{s}) \propto (\sqrt{s}R)^d$. This large phase-space
of the gravitons is in the effective description responsible for the importance of the effect. Consider e.g. the process $e^+e^- \to G + \gamma$. With the estimation for the total cross-section\footnote{Here, $G$ denotes the graviton.}
\begin{eqnarray*}  
\sigma(e^+e^- \to G + \gamma) \propto \frac{\alpha}{m_{\rm p}^2} N(\sqrt{s}) = \frac{\alpha}{s} \left(\frac{\sqrt{s}}{M_{\rm f}}\right)^{d+2} \quad,
\end{eqnarray*}
where we have used eq. (\ref{master}), we see, that the graviton processes get as important as the SM processes at energies $\sqrt{s}\approx M_{\rm f}$.

The primary observable effect for real graviton production is an apparent non-conservation of momentum on our brane, since the graviton leaves our brane and is not detected. Therefore, production of a single jet at high transverse momentum is a promising signal to look for, the main contribution for {\sc LHC} arises from the subprocess $qg \to qG$. 

Analysis of present data  yields contraints on real and virtual graviton processes and so gives a lower bound on the new scale $M_{\rm f}$ in the range $1$~ TeV, the exact value depending on $d$. For more details see e.g. \cite{revgrav} and references therein. 

\section{Black Holes}

In general relativity a (non-charged, non-rotating) black hole of mass $M$ is described by the Schwarzschild-metric. This metric is diagnonal, spherical symmetric and
has $g_{tt} = - 1/ g_{rr} = 1+ 2 \phi$, with $\phi $ the Newtonial potential. It can
be shown, that this holds too (up to factors of order 1) for the higher dimensional Schwarzschild-solution \cite{my}. Here, of course we have to use
the $d+3$ dimensional potential. That is, the zero $R_H$ of the metric coefficients, which gives the horizon radius is (again up to factors of order 1)
\begin{eqnarray*}  
 \frac{M}{M_{\rm f}^{d+2}} \frac{1}{R_H^{d+1}} \approx 1 \quad.
\end{eqnarray*}
It is not surprising, that a black hole of mass $\approx M_{\rm f}$ has  a horizon-radius of $R_H \approx L_{\rm f}$. Due to the lowered Planck-scale this horizon-radius is
in the range of distances that can be reached at {\sc LHC}-energies. Note, that we can
use the spherical symmetric solution since the black-hole radius for collider-energies is much smaller than the radius of the extra dimensions and the periodic boundary conditions can be neglected.

From general relativistic arguments\footnote{This is known as Thorn's hoop conjecture.}, the cross-section for black hole production is $\sigma=\pi R_H^2$. It has been discussed whether this cross-section is
 reliable and it has been shown in several approaches that it holds at least up
to energies $\approx 10 M_{\rm f}$ \cite{cross-section}.

Using the {\sc QCD} parton-distribution functions it is then possible to compute differential and total cross-sections for the black-hole production at {\sc LHC}. The total number of black holes depends only weakly on $d$ and is of
order $10^9$ per year, that is $\approx 30$ black holes per second \cite{BHS}!

The produced black holes will undergo evaporation which is more a decay because of their high temperature \cite{Harms}. Unfortunately, the precise description of this process falls into the regime of quantum gravity and is unknown, the main question being whether the black hole evaporates completely or whether a stable relic is left \cite{relics}. 

There are several observables for the black hole detection. First, there will be a sharp drop in the jet-spectrum at high transverse momentum $>M_{\rm f}$. High energetic jets can not be produced any more since their energy will create a black hole. The black hole decay then yields multi-jet events with energies $<M_{\rm f}$ or, in the scenario with relics resp., mono-jets. Since the black hole radiates thermally, this processes are flavour-blind. The detection of black holes would not only allow us to test the large extra dimension model but it would be an enormous exciting possibility to examine the properties of an truly
extreme state of matter on the junction between general relativity, thermodynamics and quantum field theory. 

For more details on this subject see e.g. \cite{BHS}.

\section{Minmal Length}

Even if a full description of quantum gravity is not yet available, there
are some general features that seem to go hand in hand with all promising candidates
for such a theory. Besides the the need for a higher dimensional space-time,
there is the existence of a minimal length scale. 

In perturbative string theory,
the feature of a fundamental minimal length scale arises from the fact that strings can not probe 
distances smaller than the string scale. If the energy of a string reaches the Planc-mass 
$m_{\rm p}$, excitations of the string can occur and cause a non-zero extension
\cite{stringmini}. 
Due to this, uncertainty in position measurement can never become smaller than 
$l_{\rm p}$. This can also be understood in a heuristic way. Consider an experiment to test a spacetime-structure from about Planck-length. The Compton-wavelength a particle must have in order to resolve Planck-length is just Planck-mass as follows from the uncertainty principle. Since the 
particle has Planck-mass its perturbation of the spacetime-metric can not be
neglected any longer and thus causes an additional uncertainty at high energies.
For a review, see \cite{reviewmini}.

In order to implement the notion of a minimal length into the model of {\sc LXD}s, 
let us now suppose that one can increase the momentum $p$ arbitrarily, but that the wave-vector $k$ 
has an upper bound. 
This effect will show up when $p$ approaches a certain scale $M_{\rm f}$. 
The physical interpretation of this is that particles can not possess arbitrarily
small Compton-wavelengths $\lambda = 2\pi/k$ and that arbitrarily small scales
can not be resolved anymore. 

To incorporate this behaviour, we assume a relation $k=k(p)$ between $p$ and $k$ 
which is an uneven function (because of parity) and which asymptotically
approaches $1/L_{\rm f}$.\footnote{Note that this is similar to introducing an energy dependence of Planck's
constant $\hbar \to \hbar(p)$.} There are several approaches how to deal with this generalisation, see e.g. \cite{miniapp}. We will use the analysis of \cite{ownmini}. The modified commutator algebra in quantum mechanics 
then reads
\begin{eqnarray}
[\,\hat{x},\hat{p}]&=& + {\rm i} \frac{\partial p}{\partial k} \Rightarrow \Delta p \Delta x \geq \frac{1}{2}  \Bigg| \left\langle \frac{\partial p}{\partial k} \right\rangle \Bigg| \quad. 
\end{eqnarray} 
which results in a generalised uncertainty relation. In a power series expansion of the function $k(p)$, we find
an additional term in the uncertainty, which is quadratic in $p$. A further consequence of the existence of the minimal lenght is a dropping of the 
momentum space measure with the functional determinante $\partial k / \partial p$.

This generalized uncertainty relation influences observables of high energy physics. Mainly, it states that effects
at high energies are much less then expected from the SM only. Consider e.g. the black hole formation:
the beam energy needed with the generalized uncertainty to focus enough energy-density in a small region
of space-time will increase. We want to point out that the
minimal length effects do strongly modify predictions of the {\sc LXD}-scenario and have to be included for
completeness.

\section{Summary}

{\sl ''Truth in science can best be defined as the working hypothesis best suited to open the way to the next better one.'' 
\hspace{3cm} ---  Konrad Lorenz}
\vspace{0.3cm}

The models with large extra dimensions do not claim to be a theory of everything but provide a useful basis to make predictions beyond the standard model. Observables, like graviton or black-hole production, can be used to test general
features of our spacetime, such as the number and the size of the extra dimensions or the existence of a minimal length scale.


\begin{acknowledgments}
This work was supported by {\sc GSI, DFG, BMBF} and {\sc DESY}. The authors thank all members of the {\sc LXD}-Group at Frankfurt University for fruitful discussions. We want to apologize for every contribution on the topic of Large Extra Dimension that has not been mentioned here. 
\end{acknowledgments}

\end{document}